\documentstyle[12pt]{article}

\def\be{\begin{equation}}
\def\ee{\end{equation}}
\def\bea{\begin{eqnarray}}
\def\eea{\end{eqnarray}}

\def\beq{\begin{equation}}   \def\eeq{\end{equation}}
\def\bea{\begin{eqnarray}}   \def\eea{\end{eqnarray}}

\newcommand{\matel}[3]{\langle #1|#2|#3\rangle}

\begin{document}
\begin{flushright}
UND-HEP-03-BIG\hspace*{.2em}05\\
hep-ph/0401003\\
\end{flushright}


\vspace{.3cm}
\begin{center} \Large 
{\bf 
{Is Super-$B$ Sufficiently Superb?  -- On the Motivation for a Super-$B$  
Factory}}
\\
\end{center}
\vspace*{.3cm}
\begin{center} {\Large 
I. I. Bigi$^a$ and A.I. Sanda$^b$ }\\ 
\vspace{.4cm}
{\normalsize 
$^a${\it Physics Dept.,
Univ. of Notre Dame du
Lac, Notre Dame, IN 46556, U.S.A.} \\
$^b${\it Physics Dept., Nagoya University, Nagoya 464-01, Japan}}
\\
\vspace{.3cm}
{\it e-mail addresses: bigi.1@nd.edu, sanda@eken.phys.nagoya-u.ac.jp } 
\vspace*{0.4cm}

{\Large{\bf Abstract}}

\end{center}
Despite the great success of the $\Upsilon (4S)$ $B$ factories at KEK and SLAC 
and the guaranteed addition of high sensitivity measurements on beauty decays to be performed at the Tevatron and LHC, a strong case can be made for an $e^+e^-$ Super-$B$ factory  yielding  data samples of order  
$10^{10}$ $B \bar B$ pairs as a necessity rather than luxury. It has to be justified through its ability  to not only 
establish deviations from the Standard Model, but also diagnose and interpret those in terms of 
specific features of the New Dynamics. The role to be played by a Super-$B$ factory is thus analogous {\em and even in parallel} to that of a linear collider. The latter's goal is to provide more detailed information on previously discovered New Physics involved in the electroweak phase transition. 
Likewise a Super-$B$ factory would provide precision probes for analyzing whether such New Dynamics has an impact on heavy flavour dynamics -- a need particularly manifest if the New Physics is housed under the `big tent' of SUSY.  

The huge statistics of a Super-$B$ factory and the comprehensive body of accurate measurements uniquely possible there would be harnessed in several classes of studies, among them more precise extractions of $V(ub)/V(cb)$ and $V(td)/V(ts)$, more detailed analyses of 
$B\to \gamma X_{s,d}, \, l^+l^-X$ and novel data on $B\to \tau \nu , \, \tau \nu D,\, \nu \bar \nu X$ and maybe even on $\Upsilon (5S) \to B_s \bar B_s$ -- all in addition to a host of CP asymmetries.

\tableofcontents 
\section{Executive Summary}
\label{INTRO}

\subsection{A brief look back}

The study of heavy flavour hadrons -- starting with kaons and hyperons -- has lead to many 
discoveries that were crucial for the evolution of today's Standard Model (SM). 
To cite but a few of the most seminal ones: 
\begin{itemize}
\item 
The $\tau$ - $\theta$ puzzle in kaon decays provided the first suggestion that parity is not conserved in nature. 
\item 
The observation that the production rate of some `strange' hadrons exceeded their decay rate by many orders of magnitude was explained through postulating a new quantum number -- `strangeness' -- 
conserved by the strong, though not the weak forces 
\cite{STRANGE}. This was the beginning of the second quark family. 
\item 
The weak decays of pions, kaons and muons were related through 
Cabibbo universality \cite{CABIBBO}.  
\item 
Flavour oscillations were predicted for the $K^0 - \bar K^0$ complex \cite{OSCIL}. 
\item 
The absence of flavour-changing neutral currents -- first noticed in $K_L \to \mu^+\mu^-$ and 
$\Delta M_K$ -- was implemented by introducing another quantum number `charm', which completed the second quark family \cite{GIM}. Its mass was predicted to be roughly about 2 GeV \cite{LGR}. 
\item 
CP violation -- observed through $K_L \to \pi^+\pi^-$ \cite{CRONIN} -- 
led to the postulation of yet another, the third family \cite{KM}. 
\end{itemize}
All of these features, which are now essential pillars of the SM, were New Physics at {\em that} time! 
They came as quite a surprise  -- even as a shock.   Yet later, sometimes much later, they have been confirmed, sometimes overcoming considerable skepticism in the community: 
\begin{itemize}
\item 
Charm hadrons were indeed found in the mass region around 2 GeV with the expected lifetimes of 
$10^{-13} - 10^{-12}$ sec and a preferred coupling to strange hadrons. 
\item 
Beauty hadrons and top quarks were found, mostly with the expected properties. 
The lifetimes of beauty hadrons actually turned out to be considerably longer than had been anticipated based  on a naive analogy with the Cabibbo angle. This lead to the realization that the 
CKM matrix is highly symmetrical and hierachical. We have not digested yet the message that is encoded in this peculiar pattern. 
\item 
$B_d - \bar B_d$ oscillations were found. 
\item 
Most triumphantly CP violation has been firmly established \cite{BELLECP,BABARCP} in $B_d \to J/\psi K_S$ in impressive 
quantitative agreement with CKM predictions \cite{SANDA1,SANDA2}. 

\end{itemize}
While these last items constitute impressive and novel confirmations of the SM, there is no reason to think we have reached `the end of the road' here -- on the contrary!

\subsection{From $B$ to Super-$B$ factories}

There can be no argument that the two $B$ factories at KEK and SLAC have 
already achieved great success both on the technical level -- the instantaneous as well as 
integrated luminosity -- and as far as physics results are concerned -- 
the first CP violation established outside $K_L$ decays  \cite{BELLECP,BABARCP}. 
This success can most concisely be expressed through the data on the time-dependent CP asymmetry in 
$B_d \to [\bar cc] K_S$, where $[\bar cc]$ denotes the various charmonia states: $J/\psi, \psi^{\prime}$ etc. 
as they existed in the summer of 2003, a mere four years after data taking began 
\cite{BROWDER}: 
\bea 
{\rm sin}2\phi_1 = 
 \left\{ 
\begin{array}{lll} 
0.733\pm 0.057 \pm 0.028 & {\rm BELLE}  & 140\; fb^{-1} \\
 0.741 \pm 0.067 \pm 0.03 & {\rm BABAR} & 78\; fb^{-1} \\
0.736 \pm 0.049 & {\rm  world  \; average} & 
\end{array}
\right. 
\label{PHI1WA}
\eea
in complete agreement with CKM predictions \cite{CKMFITTER}
\footnote{The angle $\phi_1$ of the CKM unitarity triangle is occasionally denoted by $\beta$.}.

This has lead to the discussion of 
so-called Super-$B$ factories, i.e.  asymmetric $e^+e^-$ colliders operating near the 
$\Upsilon(4S)$ resonance with a luminosity of about two orders of magnitude higher, namely 
a few $\times 10^{35} cm^{-2}s^{-1}$ producing samples of a few $ab^{-1}$, i.e. few 
$\times 10^9$ $B\bar B$ pairs per year. Compared 
with the anticipated yields at hadronic colliders -- about 
$4\times 10^{11}$ and $20 \times 10^{11}$ beauty hadrons per year at BTeV and LHC-b, respectively -- one would obtain a highly competitive sample of $B$ mesons produced in a 
clean environment, in particular when including detection efficiencies etc.  
While the very success of the $B$ factories and the expected high quality measurements at the 
hadronic collider makes people doubt the need for such a Super-$B$ 
factory, it is our judgment that a very strong case can be made for it. However it has 
to be justified on different grounds than for the original $B$ factories. Those are second generation facilities (following a pre-historic and first generation period), whose motivation was actually quite straightforward (although it was not always perceived that way, as we know from our own frustrating experience). At the time of their approval there were 
recognizable "killer applications": the CKM description of CP violation predicted with {\em no plausible} deniability that modes like $B_d \to \psi K_S$, $\pi^+\pi^-$ and $B^{\pm} \to D^{neut}K^{\pm}$ had to exhibit 
large CP asymmetries, some of which were predicted with high {\em parametric} reliability. This was true at a time when $\epsilon_K \neq 0$ was the only known CP violation. Altogether a realistic goal was a (semi)quantitative exploration of the landscape of heavy flavour dynamics, which 
still contained large patches of `virgin' territory. The results obtained already have promoted 
the KM paradigm from an ansatz to a {\em tested} theory. Ten years later the situation is quite different for a third generation facility -- partly due to the very success of the $B$ factories. 

\subsection{Goals for a Super-$B$ Factory}

The main argument is not so much whether the motivation for a Super-$B$ factory is superb -- most 
students of the field would probably agree -- but whether it is {\em sufficiently} superb  justifying 
its construction in a time of limited resources.  

A core goal of the $B$ factories was and still is to uncover the intervention of New Physics. 
The three family structure of the SM with its mass hierarchy and the very peculiar pattern in the CKM matrix 
undoubtedly point to the existence of physics beyond the SM -- alas in a so far obscure 
direction; we have no clear idea even about the scale that characterizes this kind of New Physics. On the other hand the strong conviction that New Physics connected with the electroweak phase transition 
has to exist at the TeV scale -- i.e. "nearby" -- is driving the construction of the LHC 
\footnote{The TEVATRON might actually catch the first glimpses of it.}.  

The most popular candidates are  supersymmetry, extra space dimensions and 
technicolour. {\em A priori} dynamics characterized by the TeV scale should have an impact on beauty (and possibly charm and $\tau$) decays. It is then imperative to make a dedicated effort to find out whether and to which degree this is the case. Such studies have to be pushed to the highest attainable level of accuracy, since large deviations 
from SM predictions might well be the exception rather than the rule; if it turns out that New Physics has no discernible impact in heavy flavour dynamics, then this would be a frustrating, yet still highly significant lesson. Supersymmetry is a case in point (and likewise for theories with extra dimensions), since it represents an intriguing {\em organizing principle} rather than a theory; i.e., 
it provides a huge tent for many classes of theories. We have made little headway in understanding its breaking (apart from ruling out various cases). Finding out from experiment whether SUSY breaking is flavour specific or non-specific is thus essential information. 
Existing constraints 
from heavy flavour transitions already tell us that SUSY has to be realized in an  
{\em a priori} very unlikely corner of its parameter space (if at all, of course) \cite{SUSY1}. 
A more detailed exploration of beauty 
(\& charm) decays can shed light onto the dynamics underlying this peculiar pattern 
and {\em thus is an indispensable part of the exploration of the electroweak phase transition}. For the latter 
leads to masses also for states other than the gauge bosons, like the fermions. This includes the {\em off-}diagonal matrix elements in the quark mass matrices  that give rise to the CKM matrix, which in turn shapes heavy flavour decays. The electroweak phase transition 
thus leads to a huge increase in the number of observables for 
flavour-{\em non}diagonal processes. This is likely to be true for New Physics in general. It holds most definitely for SUSY: to gain access to the plethora of dynamical information encoded in the 
{\em off-}diagonal elements in the squark etc. mass matrices, one has to analyze heavy flavour decays in a dedicated and comprehensive manner. 

The importance of further studies of $B$ dynamics has been recognized as illustrated by the continuing operation of the two $B$ factories and by having even two specialized detectors for running at the LHC and the TEVATRON. The relevant question is whether the measurements to be undertaken there together with data sets of about 0.5 $ab^{-1}$ from BABAR and BELLE each 
suffice to exhaust the discovery potential.  We submit that this is not the case, that data sets of 
$e^+e^- \to B \bar B$ much larger than 1 $ab^{-1}$ are needed to complement those from 
LHC-b and BTeV. This is based on two expectations: 
\begin{itemize}
\item 
 Sizable (let alone large) deviations from the SM prediction might well be the exception rather than the rule. 
 \item 
Merely  continuing the search for New Physics might be viewed as an insufficient justification for a Super-$B$ factory. There the goal has to be to shape such a facility into a {\em precision} tool that has the potential not only to reveal New Physics, but also to 
{\em diagnose its salient features}.  Those presumably less than massive deviations can be established only if the 
{\em predictions} for and {\em interpretations} of the relevant observables can be made reliable and quite precise. 
\end{itemize}
$B$ [ and charm] decays are employed as high sensitivity probes of New Physics of any kind; 
finding that New Physics is intrinsically connected to 
the family structure of the SM would be an added bonus.

Measurements at a Super-$B$ factory are uniquely able to provide us with the necessary information making it an essential element in our quest for understanding the anticipated New Dynamics. In that context we view the role of a Super-$B$ factory like that of a daughter-in-law of the LHC somewhat in parallel to that of the latter's  more free-spending daughter, namely the Linear Collider. The LHC will provide the sweeping overviews of the land of New Physics; the essential detailed mapping out of this new territory will happen at the Linear Collider concerning the electroweak symmetry breaking -- and at a Super-$B$ factory with respect to the heavy flavour puzzle of the SM model. {\em Comprehensive precision} measurements will be the main order of the day for experiments at these two facilities (although on a numerically different level); the strategies at both will be shaped by the findings at the LHC concerning the scale and type of the anticipated New Physics.

\subsection{Assets of a Super-$B$ Factory}

A Super-$B$ factory would have several assets that make it well-equipped to pursue such goals. 
The very high statistics that could be accumulated at a Super-$B$ factory in a relatively low background 
environment can be harnessed in different ways:   
\begin{itemize}
\item 
One can achieve more accuracy in extracting the CKM parameters $|V(ub)/V(cb)|$ and 
$|V(td)/V(ts)|$ from inclusive semileptonic and radiative decays, respectively, -- quite possibly to the few percent level --, which would sharpen significantly the SM predictions for the angles in the CKM unitarity triangle; those in turn can be inferred independently through measurements of various CP asymmetries that need not be done at a Super-$B$ factory. More precise SM predictions imply more sensitivity to New Physics. 
Similarly a Super-$B$ factory provides the best stage to perform high accuracy Dalitz plots studies 
involving even multi-neutral final states for $B \to 3\pi$, $4\pi$ and $3K$, that are essential in 
properly {\em interpreting} measured CP asymmetries in those channels in terms of the fundamental parameters of the theory.  
\item  
One will be able to analyze more detailed features of rare decays like $B\to l^+l^-X$, like the 
lepton spectra and their asymmetries. 
\item 
It will allow the study of novel modes $B\to \tau \nu$, $B\to \tau \nu D$ and $B\to \nu \bar \nu X$.    
\item 
It might possibly enable us to explore the new territory of $\Upsilon (5S) \to B_s \bar B_s$. 
\end{itemize}
We are not sure whether the concept of a "killer application" is a very useful one as a 
central motivator for Super-$B$. The CP asymmetry in $B_d \to \phi K_S$ that has emerged in particular in the 
BELLE data appears in stark contrast to KM predictions \cite{GROSSMAN,SANDAPHIKS}: 
\beq 
{\rm sin}2\phi_1^{eff} (B\to \phi K_S) = \left\{ 
\begin{array}{lll} 
+0.45 \pm 0.43 \pm 0.07 & {\rm BABAR}  & 110\; fb^{-1} \\
- 0.96 \pm 0.5 ^{+0.07}_{-0.11} & {\rm BELLE} & 140\; fb^{-1}
\end{array}
\right. 
\eeq
While BABAR's number is consistent with the KM expectation as expressed through Eq.(\ref{PHI1WA}), 
BELLE's result is marginally inconsistent. 
If the average of BELLE's and BABAR's numbers reflects the true value within one sigma or so, then we are facing a quite massive conflict with the Standard Model (SM); in that case we would not need Super-$B$ to establish the effect: BELLE and BABAR together 
with LHC-b and BTeV would achieve that exciting goal. 
The point is {\em not} that analyzing this mode in great detail would not be an important task -- it will be in any case. However 
we would be ill-advised to stake the case for Super-$B$ on one or two transitions -- the 
motivation has to be derived from an attractive comprehensive program. We believe that $B$ physics can still be compared to a gold mine though one that is being heavily mined. 

In presenting the case for a Super-$B$ factory, which is complementary to LHC-b, BTeV and 
the linear collider,  we will allow for theorists to get `smarter', i.e. develop even more powerful treatments; this expectation is based on the experience of the last ten years that the challenge of ever more accurate and comprehensive data inspires and helps theorists 
to refine their tools and make more observables theoretically `clean'. We will show little restraint in raising questions, yet considerable reluctance in giving answers. 

Our discussion of the future of $B$ physics will be guided by the question: "Can we answer the 
1\% challenge?" By that we mean the following: some observables will be measured with an 
experimental accuracy around the few percent level. What is it that we need to first 
{\em predict} such observables, then {\em interpret} the experimental findings and finally  
{\em diagnose} the lessons on the underlying dynamics with a commensurate accuracy?  

\subsection{Organization of the Supporting Material}

The remainder of this paper contains material explaining and supporting the statements made in this {\em Executive Summary}. It is organized as follows: in Sect.\ref{STATUS} we sketch 
what we anticipate 
the status of heavy flavour physics to be in 2010 without a Super-$B$ factory; in Sect.\ref{ACCU1}  we consider the case for accuracy with respect to extracting CKM parameters and analysing in detail transitions $B\to l^+l^-X$, $B\to \gamma X$ and in 
Sect.\ref {ACCU2} the case for measuring new decays $B\to \nu \bar \nu X$, 
$B^+\to l^+\nu$ and $B\to \tau \nu D$; in Sect.\ref{5S} we refer to the option of running on the 
$\Upsilon(5S)$ and in Sect.\ref{TAUCHARM} we list secondary motivations based on making the 
`ultimate' measurements in charm and $\tau$ physics before presenting our conclusions in 
Sect.\ref{OUTLOOK}; the resulting decision matrix will be a rather complex one.

\section{The Status Anticipated for 2010}
\label{STATUS}

Our personal crystal ball shows the following landscape of $B$ physics for 2010, i.e. after the $B$ factories 
have accumulated a total of about 1 $ab^{-1}$ in integrated luminosity and TEVATRON and LHC experiments have been operating for some time. 
\begin{itemize}
\item 
The uncertainty on $|V(cb)|$ has been reduced to the 1\% accuracy level and on 
$V(ub)$ as well as $V(td)$ to the about 10\% level with meaningful estimates of the theoretical uncertainties. 
\item 
$\Gamma (B\to \gamma X_s)$ has been well measured including the photon spectrum extending 
to energies somewhat below 2 GeV, the relevance of which will be explained later. 
$\Gamma (B \to l^+l^-X_s)$ has been measured both inclusively and exclusively on the 
10\% level. 
\item 
$B\to \mu^+\mu^-$ has been searched for down to branching ratios of about $10^{-9}$, where 
a SM signal could show up in $B_s$ decays. 
\item 
The CP asymmetry in $B_d\to J/\psi K_S$ has been measured to within a percent or so; 
CP asymmetries in $B_d \to \pi^+ \pi ^-$, $3\pi$, $\phi K_S$ have been measured or probed with 
similar experimental sensitivity and likewise for $B\to K \pi$ etc. 
\item 
$B_s - \bar B_s$ oscillations have been found and well measured. This might actually turn out to be the least certain part of our expectations. 
\item 
CP violation has been searched for in $B_s \to J/\psi \phi$, $J/\psi \eta$ as a window onto New Physics. 
\end{itemize}
Clearly a large amount of valuable information will have been gathered then. Before evaluating it 
let us consider the `big picture' of high energy physics. We see three possible scenarios to emerge in the next several years. 
\begin{enumerate} 
\item 
{\em The optimal scenario}: New Physics has been observed in "high $p_{\perp}$ physics", i.e. through the production of new quanta at the TEVATRON and/or LHC. Then it is {\em imperative} to study the impact of such New Physics on flavour dynamics; even if it should turn out to have none, this is an important piece of information. Knowing the typical mass scale of that New Physics from collider data will be of great help to estimate its impact on heavy flavour transitions.  

\item 
{\em The intriguing scenario}: Deviations from the SM have been established in heavy flavour decays -- like the asymmetry in $B \to \phi K_S$ -- without a clear signal for New Physics in high $p_{\perp}$ physics. 

\item 
{\em The frustrating scenario}: No deviation from SM predictions have been identified. 

\end{enumerate}
We are optimistic that it will be the `optimal' scenario, quite possibly with some elements of the 'intriguing' one. Of course one cannot rule out the `frustrating' scenario; yet we would not treat it as a case for defeatism: a possible failure to identify New Physics in future experiments at the hadronic colliders (or the $B$ factories) does not -- in our judgment -- invalidate the persuasiveness of the theoretical arguments pointing to the incompleteness of the SM. It should actually be seen as a call to extend our lines of attack, one of which had to be a Super-$B$ factory. 

To conclude this crystal ball glazing: 
\begin{itemize}
\item 
Many CP asymmetries and rare decays will be measured quite accurately even without a Super-$B$ factory. As a matter of fact, the experimental accuracy on CP asymmetries like in $B_d \to J/\psi K_S$, $D \bar D$, 
$\pi^+\pi^-$, $\pi^0\pi^+\pi^-$, $K_SK^+K^-$, $B\to K\pi$, $B_s \to J/\psi \phi$ etc. and rare decays 
like $B_{d,s} \to \mu^+\mu^-$, $B\to l^+l^- K^{(*)}$ that will be achieved at the TEVATRON and LHC will not be surpassed by data from a Super-$B$ factory. 
\item 
One limitation in the accuracy of the SM predictions is due to the uncertainties in the values of 
$|V(ub)/V(cb)|$ and $|V(td)/V(ts)|$. 
\item 
Likewise there is a relevant limitation in our ability to relate an observed CP asymmetry to the CKM parameters due to hadronic uncertainties, often very unkindly referred to as `Penguin pollution'. 
\end{itemize}

Two related questions then arise: (i) To which degree is the exploitation of such experimental sensitivity limited by theoretical uncertainties; how can those be reduced? (ii) More generally, can BABAR/BELLE/LHC-b/BTeV [\& CLEO-c and BESIII] exhaust the discovery potential in beauty [\& charm] decays? 

Obviously a breakthrough in theoretical technologies would be of great help, but we cannot count on it to happen. Yet more data of higher quality can provide considerable assistance in reducing 
theoretical uncertainties. For while it is true that there are more parameters controlling beauty decays in a significant way than, say, it is the case for strange decays, the number of observables is much higher still. I.e., there are many correlations among them, the best 
known one expressed through the KM unitarity triangle. Determining the same parameter in more than one way will thus be a powerful element of `quality control' both on the experimental and theoretical side. 

The goals can be grouped as follows: 
\begin{enumerate}
\item 
One can refine the SM predictions for the angles in the CKM unitarity triangle  
by extracting more precise values for $|V(ub)/V(cb)|$ and $|V(td)/V(ts)|$.  A Super-$B$ factory 
provides unique access to theoretically clean observables. 
\item 
One wants to interpret more reliably measured CP asymmetries in terms of the microscopic SM (or New Physics) parameters. For example one can determine the angle $\phi_2/\alpha$ from CP asymmetries measured in different $B_d \to \rho \pi$ channels. Yet their amplitudes have to be 
extracted from a detailed Dalitz plot analysis of $B_d \to 3\pi$; even if $B_d \to \rho \pi$ dominates 
$B_d \to 3\pi$, one {\em cannot} ignore other contributions. Comparing 
$B_d \to \pi^+\pi^- \pi^0$ with  $B_d \to 3 \pi^0$ and 
$B^{\pm} \to \pi^{\pm}\pi^+\pi^-, \, \pi^{\pm}\pi^0\pi^0$ will be of great help in disentangling hadronic complexities. 
\item 
$B$ decays leading to $\tau$ leptons are a valuable tool for cross checking theoretical control and a sensitive probe of New Physics. 
\item 
The polarization of the photon in radiative $B$ decays is another observable highly sensitive to the presence of New Physics. 
\item 
The very rare transitions $B\to l^+l^-X$ provide excellent probes for New Physics and its features. 
The theoretical control over {\em inclusive} rates is highest. 
\item 
Decays $B\to \nu \bar \nu X$ have great potential to provide complementary information on New Physics. 
\end{enumerate}
While experiments at the TEVATRON and the LHC, in particular LHC-b and BTeV, will 
study profitably item 5. through exclusive modes, items 2. (more than one neutral in the final state), 3. 4., 6. appear clearly to be beyond their  reach; they will contribute significantly to item 1. through analysing exclusive $B$ decays and searching for $B_s - \bar B_s$ oscillations, yet not with the breadth and cross checks available at a Super-$B$ factory, as explained later. With respect to the BABAR and BELLE experiments a Super-$B$ factory holds out the promise of  harnessing its much larger statistics to make these measurements considerably more precise, provide more cross checks to evaluate theoretical uncertainties and transform `mere' observations of a rare transition into a true measurement of it. 
The huge luminosity anticipated for a Super-$B$ factory can be employed to enhance the 
experimental control even further by producing intense  monoenergetic beams of $B$ mesons  
through fully reconstructing one of the $B$ mesons in $e^+e^- \to B \bar B$. This  
enables us to perform measurements of high precision and sensitivity in a highly controlled environment. Again this will allow {\em novel} rather than merely new measurements.

As indicated above with the information available from the $B$ factories the uncertainty on 
$|V(cb)|$ will be reduced to the 1\% level;  as sketched later, with data from Super-$B$ factories 
in hand one can reduce the uncertainty on $|V(ub)|$ and $|V(td)|$ down to the 10\% and quite possibly to the 5\% level. 
{\em In our judgment such accuracy levels are a clear necessity rather than a luxury to make full use of the discovery potential for New Physics in beauty decays}. It is quite 
conceivable that New Physics impacts CP asymmetries in $B\to J/\psi K_S$, $D\bar D$, 
$\pi\pi/\pi\pi\pi$, $DK$, $K\pi$ etc. `massively' shifting them away from their SM values by 
{\em several} $\times 10\%$, even changing their signs. Yet the SM with the CKM implementation of CP violation has so far provided a successful description of strange, charm and beauty decays with observables covering about eight orders of magnitude in energy scale; a priori it appears unlikely that New Physics could have remained hidden unless its contributions are generically smallish relative to the SM (or it is intrinsically connected to the quark flavour structure). Therefore  one cannot {\em count} on New Physics suddenly making such a dramatic and obvious appearance in $B$ decays; deviations from SM predictions of no more than 10 percentage points -- say from 40\% to 50\% -- could actually be on the large side. Of course there are special cases where the SM prediction is suppressed due to very specific reasons of the SM like the CP asymmetry in 
$B_s\to J/\psi \phi, \, J/\psi \eta$ \cite{SANDA2}, which cannot exceed about 2 \%. Since one cannot {\em count} on large deviations, it is mandatory to aim for the highest achievable accuracy in measurements and interpretations. Finally huge statistics are required, since it is important to measure rate transitions rather than `merely' observe their existence.

We will illustrate the case for a Super-$B$ factory by four classes of case studies: 
\begin{enumerate}
\item 
Extracting $|V(ub)|$ and $|V(td)|$ in addition to $|V(cb)|$; 
\item 
measuring $B\to \mu \nu$, $\tau \nu$ and $\tau \nu D$; 
\item 
studying CP violation in $B\to 3\pi$, $3K$; 
\item 
analyzing $B\to \gamma X_q$, $l^+l^-X_q$ and $\nu \bar \nu X$ including their CP asymmetries.

\end{enumerate}

\section{The Case for Accuracy}
\label{ACCU1}

\subsection{Part I: CKM parameters}

\subsubsection{$V(cb)$}
\label{VCB}
There are three methods for extracting $|V(cb)|$ that invoke QCD in a credible way, namely from 
(i) $\Gamma_{SL}(B)$ -- the `golden' way --, (ii) $B\to l \nu D^*$ at zero recoil -- the 
`gold-plated' way -- and possibly (iii) $B \to l \nu D$ -- the `Cinderella story'. The first two are being used extensively with the `golden' way yielding in the near future a value within 1-2 \% that can be checked by the `gold-plated' method within the latter's larger uncertainties. 

For the subsequent discussion a few general remarks on the Heavy Quark Expansion (HQE) and 
its Heavy Quark Parameters (HQP) are appropriate. The HQE allows to describe rates for a host of inclusive transitions -- $b \to c$ and $b\to u$ semileptonic and radiative ones -- through HQP, namely expectation values of a {\em universal} cast of local quark-gluon operators. Those HQP can be extracted from the shape of an energy or mass distribution -- say in $B\to l \nu X_c$. The shape information is concisely encoded in that distribution's {\em moments} 
of different orders. In general there is {\em not} a one-to-one correspondence between these HQP and the moments; i.e., the former are obtained from nontrivial 
linear combinations of the latter. The fact that the HQP can be determined from different types of moments, namely leptonic, hadronic or 
photonic moments allows to greatly overconstrain them, which  provides a high degree of quality control over systematics on the theoretical as well as experimental side. 
{\em Once the HQP are obtained from moments of $B \to l \nu X_c$ transitions, they can be used perfectly well for $B\to l \nu X_u$ and $B\to \gamma X_s$}. Claiming one needs to measure moments of $b\to u$ decays to obtain the HQP for describing them is incorrect. 

$\bullet$ {\em The Cinderella story}: A novel tool has been put forward by Uraltsev  -- the `BPS expansion' 
\cite{BPS}. 
 It starts from the observation that if the chromomagnetic and kinetic moments had identical values -- 
$\mu_G^2 \equiv \matel{B}{\bar b \frac{i}{2}\sigma \cdot Gb}{B}/2M_B = 
\mu_{\pi}^2 \equiv \matel{B}{\bar b(i\vec D)^2 b}{B}/2M_B$ -- were to hold many simplifications arise concerning the power corrections. While this scenario does not -- and could not -- hold exactly in QCD, it holds approximately, namely for $\mu_{\pi}^2 - \mu_G^2 \ll \mu_{\pi}^2$. The formfactor for $B \to D$ can then be calculated with the main challenge provided by the proper inclusion of perturbative corrections. It appears that the intrinsic theoretical uncertainty can be 
 reduced to the very few percent level meaning that one could extract $|V(cb)|$ with maybe about very few percent 
theoretical uncertainty.

$\bullet$  Since the SM $V-A$ charged weak currents are suppressed by $V(cb)\simeq 0.04$ in amplitude, New Physics in the form of charged currents of different chirality -- in particular of the $V+A$ type -- could surface in semileptonic $B$. 
 They would affect $B\to l \nu X_c$ differently -- especially the lepton spectrum -- than $B\to l \nu D^*$ or 
 $B\to l \nu D$. Finding significantly different values for $|V(cb)|_{incl}$ and $|V(cb)|_{excl}$ could thus point to an admixture of right-handed currents in the $b\to l \nu c$ transition \cite{VOLVCB}. 
 However to make such a conclusion convincing would probably require validation through similar findings in 
 $B\to \tau \nu X_c, \; \tau \nu D^*, \; \tau \nu D$; accurate data for those channels would again require  a 
 Super-$B$ factory.

\subsubsection{$V(ub)$}

A popular opinion can be summarized as follows: measure the exclusive mode 
$B\to l \nu \pi$ and -- `in the lattice we trust' -- determine $|V(ub)|$ employing the lattice QCD's evaluation of the form factor. We would like to invoke a well known quote from someone who has often been referred to as a theorist -- however being theorists ourselves we view him rather as a highly competent machine builder --, namely Lenin: "Trust is good -- control is better". Accordingly we see it as mandatory to determine $|V(ub)|$ using a very different method, namely from inclusive 
semileptonic $B \to l \nu X_u$ decays. 

The total width $\Gamma (B\to l \nu X_u)$ is calculated through the OPE in terms of the same HQP that control  
$B\to l \nu X_c$ and are extracted there -- with the exception of $m_c$, which is irrelevant for $b\to u$. 

While the first direct evidence for $V(ub) \neq 0$ came from observing charged leptons with energies beyond the kinematic limit for $b\to c$, this method does not lend itself to a precise determination due to considerable theoretical uncertainties. About 90\% of the $b\to u$ signal is buried underneath the much larger $b\to c$ signal. There is no model-independent description of the shape of the 
{\em end-point} spectrum; the relationship with the distribution function that can be measured in 
$B\to \gamma X$ has unknown $1/m_Q$ corrections. The endpoint spectrum is predicted to be 
significantly different in $B_d$ and $B_u$ decays; this difference is driven by the expectation values of four-fermion 
operators arising in order $1/m_Q^3$ \cite{WA} 
(conveniently referred to as weak annihilation). Duality limitations quite possibly could be very 
significant, when one can use only the fairly narrow endpoint region \cite{VADE}. 

A much better way is to measure the hadronic recoil mass and partially integrate it, i.e. 
\beq
\Gamma (M_X)\equiv \int ^{M_{X,max}}dM_X d\Gamma (B\to l \nu X)/dM_X  \; \; \; 
{\rm with} \; \; \;  M_{X,max} < M_D \; . 
\eeq 
For it provides the best kinematic discrimination against $b\to c$ transitions 
with only about 10\% of the $b\to u$ signal buried under the huge $b\to c$ signal. Since there 
will be "leakage" from the dominant $b\to c$ signal to $M_X < M_D$ due to measurement 
errors, one has to select an $M_X$ cut below $M_D$ by a certain margin. This of course 
introduces some theoretical uncertainty, yet only a mild one for $M_X \simeq 1.6$ GeV or even higher.  

Another source of theoretical uncertainty lies in the low $q^2$ region of the distribution with 
$q$ denoting the momentum of the lepton pair $l\nu$. Two ways have been suggested to deal with it.  
(i) One imposes a lower cut on $q^2$\cite{BAUER}. The advantage of this procedure is that it can be done 
of course. Yet there are drawbacks as well, in particular of introducing a significant dependence on higher-order 
effects, which are difficult to control. (ii) Alternatively one can infer the low-$q^2$ part of the recoil mass distribution from the photon energy spectrum in 
$B\to \gamma X$. Yet for this purpose one needs to measure the photon spectrum also below 2 GeV, at least down to 1.9 GeV or even better to 1.8 GeV \cite{SPECTRUM}, but not below. 
It is advisable to employ both approaches to insure some quality control. 

Preliminary studies \cite{TORU} indicate that determining $|V(ub)|$ with a theoretical uncertainty of no more than 5\% seems feasible through dedicated analyses of $\Gamma (M_X)$.  It might well be that a Super-$B$ factory is needed to provide the required high-quality data on semileptonic and radiative $B$ decays.

\subsubsection{$V(td)$}

The usual observables for determining $|V(td)|$ are (i) $\Gamma (K^+\to \pi^+\nu \bar \nu)$, 
(ii) $\Gamma (B\to \gamma \rho/\omega)$ vs. $\Gamma (B\to \gamma K^*)$ and 
(iii) $\Delta M(B_s)$ vs. $\Delta M(B_d)$. All three of them are driven by loop-processes within the SM and thus could be affected by New Physics. The first one is theoretically the cleanest one, yet experimentally the hardest. The second one is most vulnerable to a lack of theoretical control over long-distance dynamics. The third one is usually seen as the best candidate -- unless 
$B_s-\bar B_s$ oscillations proceed overly fast. 

In any case it seems highly worthwhile to consider other promising cases as well. There seems to have emerged considerable optimism that one could distinguish $B\to \gamma X_d$ from the 
dominant $B\to \gamma X_s$, although there are no global kinematic discriminators available. 
Yet this seems feasible only at a Super-$B$ factory. It should be possible to develop a treatment that is sufficiently clean theoretically.

\subsection{Part II: CP Violation in $B\to 3\pi$, $3K$}

Many nonleptonic $B$ decay modes are expected to show sizable or even large CP asymmetries. 
Yet the relationship 
between the observed asymmetry and the underlying weak phase is  complicated by the presence of more than 
one transition operator and hadronization effects. Theoretical schemes like "pQCD" 
\cite{PQCD} and "QCD factorization" \cite{QCDFACT} will help 
to shed light on these issues, but by themselves cannot be expected to provide precise prescriptions.   

The mode $B_d \to \rho \pi$ has been suggested \cite{SNYDER} for determining the angle $\phi_2$ a.k.a. 
$\alpha$ in the CKM unitarity triangle. However one has to extract this signal from a Dalitz plot analysis of $B \to 3\pi$, and this will require very careful work, if one wants to determine the angle without large theoretical uncertainty. For there will be other contributions even close to the $\rho$ bands. Chiral dynamics tell us that there have to be contributions of the type $\sigma \pi$ as well, and the so-called $\sigma$ resonance or enhancement is {\em not} described by the usual Breit-Wigner excitation curve \cite{ULF}. Furthermore since the $\sigma\pi$ final state carries a CP parity opposite to  that of $\rho \pi$, it will tend to contribute with the overall CP asymmetry in $B\to 3 \pi$ 
with the opposite sign and thus could have a sizeable impact even with a small branching ratio.  In addition there might well be other configurations interfering with the $\rho \pi$  state thus producing  a contribution linear in their amplitude as their main effect. Once one aims at a about 10 \% accuracy one cannot ignore such complexities. 

For these reasons I find present extrapolations about how well 
this angle can be determined given a certain integrated luminosity somewhat academic at present. It is primarily a challenge to theory to deal with these complications. Yet data on 
$B^{\pm} \to \pi^{\pm}\pi^0\pi^0$ and $B_d \to 3 \pi^0$ are bound to provide valuable information on this issue. It seems that only a Super-$B$ factory can yield such data. 

Such complications arise even more when one undertakes to extract $B\to \rho \rho$ from 
$B \to 4 \pi$ \cite{RHORHO}. 

Similar issues arise for a CP asymmetry in $B_d \to \phi K_S$: it has to be extracted from a Dalitz 
plot analysis of $B_d \to K^+K^-K_S$. Again, considerable care has to be applied in differentiating  
$B_d \to \phi K_S$ against $B_d \to f_0K_S$ (or other configurations, where $K^+K^-$ form a scalar partial wave); for with $f_0K_S$ being CP even in contrast to the CP odd 
$\phi K_S$, it will tend to contribute a CP asymmetry with the opposite sign and could thus dilute 
the CP asymmetry significantly. One should note that in the corresponding charm mode 
$D^0 \to  K^+K^-K_S$ there is a significant $f_0$ contribution underneath the $\phi$ peak. The same should hold for $B_d \to f_0K_S$ vs. $B_d \to \phi K_S$, unless the effective weak operators 
driving $D^0 \to 3 K$ and $B_d \to 3 K$ possess a different chirality structure (which would 
be natural in the presence of New Physics). Of course, such a pollution could only reduce the CP asymmetry, not enhance it. So this caveat is relevant mainly if the CP asymmetry in 
$B_d \to \phi K_S$ were observed to fall between the SM expectation and zero. To deal with it in a satisfactory way, one had to apply a full-fledged Dalitz analysis unfolding different partial waves, which again requires very high statistics.

\subsection{Part III: $B\to l^+l^-X$, $B\to \gamma X$}

A Super-$B$ factory would produce of the order of $10^6$ $B\to \gamma X$ events per year. While such a huge data sample is not necessary for a precise determination of 
$\Gamma (B \to \gamma X)$, two novel applications have been mentioned already: 
\begin{itemize}
\item 
Distinguishing $B\to \gamma X_d$ against $B\to \gamma X_s$ opens up a new avenue for determining $V(td)/V(ts)$. 
\item 
Measuring the photon spectrum below 2 GeV -- preferably down to 1.8 GeV or so -- would allow us 
to get a better extraction of the quark distribution function and thus help us in the $b\to u$ analysis.

\end{itemize}
In addition there is another observable that would be very sensitive to New Physics: 
\begin{itemize}
\item 
The SM predicts that in $b\to \gamma s$ the $s$ quark is purely left-handed. 
Accordingly the photon in $B_d,\, B^- \to \gamma X$ [$\bar B_d,\, B^+ \to \gamma X$] is 
predominantly left[right]-handed \cite{PIRJOL}. This selection rule, however, does not 
necessarily hold in the presence of New Physics. Measuring the photon polarization thus provides a sensitive probe for the presence of New Physics. The most promising way to achieve this is to measure angular correlations in exclusive $B\to K^{**}\gamma \to (K\pi \pi)\gamma$ modes 
\cite{PIRJOL}. 
\end{itemize}

The landscape becomes richer in $B\to l^+l^-X$ transitions and thus even more promising reveal New Physics. In addition to the total rate one can analyze the $l^+l^-$ spectrum, the forward-backward and the CP asymmetry in the $l^+$ vs. $l^-$ energy spectra. This increase in the number of sensitive 
observables is matched by an increase in the number of effective transition operators relative to the situation in $B\to \gamma X$. To be able to exploit this discovery potential one has to deal with the challenge to accumulate sufficient statistics for these very rare modes. There is another experimental challenge one would like to be able to deal with, namely to measure inclusive transitions as much as possible, since the theoretical control over them is substantially better than for exclusive channels. Hadronic colliders do not offer this possibility, only (Super-)$B$ factories 
(and a Giga-Z)do. 

To be more specific: within the SM one predicts \cite{ALIETAL} 
\beq 
{\rm BR}(B\to l^+l^-X) = (4.2 \pm 0.7)\cdot 10^{-6} 
\eeq
(for $M(l^+l^-) > 0.2$ GeV) 
implying a sample of few$\times 10^4$ of such events per year at a Super-$B$ factory. One has to pay a certain price in getting at these events. Accordingly one can make a good measurement of the 
total rate and meaningful studies of the lepton spectra, yet not highly precise ones. 

It has been noted that the zero of the forward-backward asymmetry in the exclusive channel 
$B \to l^+l^-K^*$ is quite model-{\em in}dependent and thus can be used at a hadronic collider to probe for New Physics in a highly meaningful way \cite{GUDRUN}.  

We conclude that a Super-$B$ factory with an integrated luminosity of 10 $ab^{-1}$ is not a luxury for these measurements -- it is a necessity.

\section{The Case for More Decays: $B^+\to l^+\nu$, $B\to \tau \nu D^{(*)}$, $B\to \nu \bar \nu X$}
\label{ACCU2}

\subsection{$B^+\to \mu^+\nu, \, \tau^+\nu$}

Within the SM one predicts 
\bea 
{\rm BR}(B^- \to \tau^- \nu) &\sim& 10^{-4}\cdot \left(\frac{f_B}{200\, {\rm MeV}}\right)^2  \\
{\rm BR}(B^- \to \mu^- \nu) &\sim& {\rm few}\times 10^{-7} \cdot \left(\frac{f_B}{200\, 
{\rm MeV}}\right)^2 
\eea

While $B^- \to \tau ^-\nu$ and $B^- \to \mu ^-\nu$ cannot be measured in hadronic collisions, it 
appears that the statistics of  a Super-$B$ factory are needed to measure 
them with decent  accuracy. Both rates are actually of interest, $B^- \to \mu ^-\nu$ to provide a handle on the size of 
$f_B$ and $B^- \to \tau ^-\nu$ to probe for Higgs dynamics.

\subsection{$B \to \tau \bar \nu D^{(*)}$}

We have already stated in Sect.\ref{VCB} that inferring inconsistent values of $V(cb)$ from $B\to l \nu X_c$, $B\to l \nu D^*$ and/or $B\to l \nu D$ could signal the presence of New Physics in semileptonic $B$ decays, like in particular 
right-handed currents.  Such an interpretation would become more conclusive if it were confirmed by similar 
discrepancies in $B\to \tau \nu X_c$ vs. $B\to \tau  \nu D^*$ vs. $B\to \tau \nu D$. Such studies presumably 
require the statistics of low-background events obtainable only at a Super-$B$ factory. 

Another probe for New Physics employs $B\to l\nu D$ by  
pursuing the following program: 
 \begin{enumerate}
 \item 
Relying on the `BPS' approximation one extracts $|V(cb)|$ from $B\to e/\mu \nu D$ and 
compare it with the `true' value obtained from $\Gamma _{SL}(B)$. 
 \item 
 If this comparison is successful -- which would constitute a validation for the theoretical control over 
 $B\to l \nu D$ -- one applies the `BPS' approximation to the mode 
 $B \to \tau \nu D$, where a second form  factor becomes measurable, since $m^2_{\tau}$ is not irrelevant on the scale $M_B^2$.  
 \item 
One compares the measured ratio $\Gamma (B\to \tau \nu D)/\Gamma (B\to \mu \nu D)$ with its SM prediction. 
Contrary to some claims in the literature the hadronic formfactors do not drop out from the ratio for finite values 
of $m_b$ and $m_c$; instead they have to be treated through the BPS expansion.
A discrepancy then implies a New Physics contribution, for which the exchange of a charged Higgs would be the most obvious candidate \cite{TANAKA}: In the `large tg$\beta$' scenarios of two-Higgs-doublet models a significant deviation 
from the SM expectation can arise for charged Higgs masses of around a few hundred GeV.

 \end{enumerate}
 
\subsection{$B \to \nu \bar \nu X$}

The inclusive rate for $B \to \nu \bar \nu X$ can reliably be calculated in the SM \cite{NUNUBAR} (and is actually 
larger than for $B\to l^+l^-X$). A sample of 
$10^{10}$ $e^+e^- \to \Upsilon(4S) \to B \bar B$ would contain about $7\cdot 10^5$ SM events. It obviously represents 
a formidable challenge to identify a $B \to \nu \bar \nu X$ event -- or even $B\to K^{(*)}\nu \bar \nu$ for that 
matter. A Super-$B$ factory holds out a realistic hope to achieve that. The only other setup with the potential for observing this transition would be a Giga-$Z$ factory \cite{GIGAZ} producing $10^9$ events $e^+e^- \to Z^0$, with the two 
$b$ jets populating distinct hemispheres.  

The justification for taking on this challenge lies in the fact that the `off-shell' radiative transitions $B\to l^+l^-X$ and 
$B\to \nu \bar \nu X$ possess considerably more sensitivity to New Physics than 
$B\to \gamma X$ since more effective operators can contribute, and by the same token 
they represent more `surgical' probes. Furthermore $B\to l^+l^-X$ and 
$B\to \nu \bar \nu X$ are complementary in diagnosing New Physics couplings with 
$l^{\pm}$ [$\nu$] representing the `down' [`up'] member of an SU(2) doublet 
\cite{NUNUBARNEW}.

\section{The Case for New Territory: $e^+e^- \to B_s \bar B_s$}
\label{5S}

The option to run at $\Upsilon (5S) \to B_s \bar B_s$ might turn out to be very valuable. 
The motivation would {\em not} be to perform measurements that can be done at LHC and the TEVATRON like searching for $B_s - \bar B_s$ oscilllations and CP asymmetries in 
$B_s(t) \to D_sK, \, J/\psi \phi$; instead one would consider four goals: 
\begin{itemize}
\item 
Using EPR \cite{EPR} correlations in the coherent process $\Upsilon (5S) \to B_s \bar B_s \to f_a f_b$ with $f_a,f_b$ denoting flavour non-specific states -- in particular CP eigenstates -- and flavour specific states one can probe CP asymmetries that involve $B_s - \bar B_s$ oscillations yet with the latter proceeding so spectacularly fast that they cannot be resolved directly \cite{BOOK,UPS}. Likewise final state phase shifts can be measured in close analogy to what has been described \cite{TAUCHARM} for 
$e^+e^- \to \psi (3770) \to D^0 \bar D^0 \to f_a f_b$ or 
$e^+e^- \to \psi (3770) \to B_d \bar B_d \to f_a f_b$. 

\item 
While the modes $B_{d,s} \to \mu^+\mu^-$ (or even $B_{d,s} \to e^+e^-$) can profitably be searched for at hadronic 
colliders, this is unlikely to be true for $B_s \to \tau^+\tau^-$. Potentially this could be a very revealing channel, if 
third family fermions possessed anomalous couplings. 

\item 
Through tagged events one can measure the absolute scale of $B_s$ branching ratios, which would provide engineering input to other studies. 
\item 
In particular one could then determine $\Gamma _{SL}(B_s)$ and 
$B_s \to l \nu D_s^*$ at zero recoil. These observables could be obtained only in this way, and from them one could then extract 
$|V(cb)|$ in close analogy to nonstrange $B$ decays. This is another example of following 
Lenin's dictum "Trust is good -- control is better!". For comparing $V(cb)$ as inferred from 
$B_d$, $B_u$ and $B_s$ decays provides a powerful check on experimental systematics, yet even more on theoretical uncertainties like the often mentioned conceivable limitations to quark-hadron duality. Such limitations could be larger than predicted due to the accidental "nearby presence" of a hadronic resonance of appropriate quantum numbers. This would be a stroke of bad luck, but could happen. Due to the isospin invariance of the strong interactions it would affect 
$B_d \to l \nu X_c$ and $B_u \to l \nu X_c$ equally, but in all likelihood {\em not} $B_s \to l \nu X_{c\bar s}$ 
\footnote{A "nearby"  resonance would impact $B_d \to l \nu X_u$ and $B_u \to l \nu X_u$ differently.}. 
Such a scenario would reveal itself by yielding inconsistent values for $V(cb)$ from 
$B$ and $B_s$ semileptonic decays.

\end{itemize}
It is understandable if BELLE and BABAR want to focus on their original menu rather than spend substantial time running on the $\Upsilon (5S)$ resonance. Yet for a Super-$B$ factory as a next generation facility it is mandatory to be prepared 
for surprises and therefore allow itself the option of substantial high-luminosity running on the $\Upsilon (5S)$ resonance.

\section{Secondary Motivations: `Definitive' Measurements in Charm \& $\tau$ Physics}
\label{TAUCHARM}

According to the SM weak dynamics charm transitions are a rather dull affair with quite slow 
$D^0 - \bar D^0$ oscillations and small CP asymmetries. This picture is fully consistent with present phenomenology: 
(i) $x_D \equiv \frac{\Delta m_D}{\Gamma_D} < 3\%$, 
$y_D\equiv \frac{\Delta \Gamma_D}{2\Gamma_D} = (1 \pm 0.5) \%$. (ii) CP asymmetries are at most a few to several percent. \cite{CICERONE}. 

Yet instead of viewing the `glass as half empty', we see it as `half full' in the sense that if one 
observes something truly interesting, one has discovered New Physics. Rather than considering it merely a wild goose chase one should note that charm transitions represent a very unique laboratory: for charm is the only up-type quark allowing the full range of indirect searches for the footprints of New Physics in oscillations and CP asymmetries \footnote{Top quarks decay before they can hadronize; without top mesons $T$ there can be $T^0 - \bar T^0$ oscillations nor can hadronization provide other forms of `cooling' to maintain coherence between different transitions 
\cite{RAPALLO}.} 

$D^0 - \bar D^0$ oscillations will occur at some level in the SM, Cabibbo suppressed decays will exhibit some direct CP violation etc. The question is at which level that will happen. 
The main motivation is to find unambiguous signs of New Physics. Accordingly we define as 
`definitive' measurements whose sensitivity reaches down to the SM level. 

$\bullet$ $D^0 - \bar D^0$ {\em oscillations}: Very conservatively one has 
$x_D$, $y_D \leq {\cal O}(1\%)$; more refined estimates yield $x_D$, $y_D \leq {\cal O}(0.1\%)$ 
\cite{DDOSC,FALK}. 
These numbers -- and the reliability with which they are obtained -- 
leave very little space for invoking the observation of $D^0 - \bar D^0$ oscillations as 
conclusive evidence for New Physics, in particular if one finds similar values for $x_D$ and $y_D$. 
Nevertheless it is very important to make dedicated efforts to measure or at least to lower 
the bounds on $x_D$ and $y_D$: for values of $x_D \sim 0.01 \sim y_D$ would create a sizeable 
{\em bias} on the value of the angle $\phi_3/\gamma$ extracted from CP asymmetries in 
$B^{\pm} \to D^{neut}K^{\pm}$ \cite{SILVA}. 

\subsection{CP Violation in $D$ Decays}

$\bullet$ {\em Time dependent CP asymmetries} can arise in 
$D^0(t) \to K^+K^-, \, \pi^+\pi^-, \, K_S\phi$ -- like for $B_d(t) \to J/\psi K_S$ -- or in the 
doubly Cabibbo suppressed mode 
$D^0(t) \to K^+ \pi^-$ -- analogous to $B_d(t) \to \rho^+ \pi^-$. For the former [latter] 
category one wants to probe down to the ${\cal O}(10^{-4}$ [ ${\cal O}(10^{-3}$],  since no SM signal can arise at those levels \cite{BSD,KPI}.  

$\bullet$ One should search for {\em direct} CP violation in the partial widths for 
$D^{\pm} \to K_{S[L]}\pi^{\pm}$ and for a host of singly Cabibbo suppressed channels down to the ${\cal O}(10^{-3})$ level \cite{YAMAMOTO}. Any direct CP asymmetry in Cabibbo allowed and doubly 
forbidden channels (except for modes like $D^{\pm} \to K_{S[L]}\pi^{\pm}$ would establish New Physics. 

$\bullet$ CP asymmetries in final state distributions -- Dalitz plots, T-odd correlations etc. -- of 
order 1 \% also would show the presence of New Physics \cite{CICERONE}. 

\subsection{$\tau$ Decays}

The recent observation of neutrino oscillations has opened another avenue for probing fundamental 
dynamics. Since they signal the presence of lepton number violating transitions, it is mandatory to search for other 
such processes -- like $\mu \to e \gamma$ or in particular $\tau \to \mu \gamma$. 
Furthermore the question of whether 
the baryon number of the universe might be due to primary lepto- rather than baryo-genesis, provides strong motivation 
to search for CP violation in the lepton sector. In particular rare $\tau$ decays hold the promise to reveal 
such effects \cite{OKADATODD}.

\section{Our Conclusions}
\label{OUTLOOK}

It is no exaggeration to state that comprehensive studies of heavy flavour dynamics 
\begin{itemize}
\item 
are of fundamental importance, 
\item 
its lessons cannot be obtained any other way and 
\item 
they cannot become obsolete!

\end{itemize}
No matter what studies of high $p_{\perp}$ physics at the TEVATRON and the LHC will or will not show -- dedicated analyses of heavy flavour transitions will remain crucial in our efforts to reveal nature's `Grand Design'. 
{\em Any direct findings of New Physics in high $p_{\perp}$ physics will actually enforce the need for and at the same time facilitate the search for further manifestations of this New Physics in flavour non-diagonal processes.}  

Yet we cannot count at all on massive manifestations in $B$ decays. Therefore we have to  address the `1\% challenge: Can we predict (certain) high value observables with 
${\cal O}(1\%)$ uncertainties, measure them, interprete the results and even diagnose the features 
of New Physics with a commensurate accuracy?  

We think such a goal is within reach rather than utopian -- if one has the tools to harness a whole network of observables and measurements of inclusive as well as exclusive reactions providing numerous cross checks. This also means we cannot afford to view lattice QCD as a panacea for all theoretical ills. Our conclusion is that a Super-$B$ 
factory yielding a sample of 
10 - 50 $ab^{-1}$ is an essential and unique element for such an admittedly ambitious undertaking. The primary justification has to come from $B$ physics -- yet charm (and $\tau$) physics provide enticing secondary motivations. 

The decision matrix for a Super-$B$ factory is necessarily a very complex one involving a host of relevant factors rather than one or two `killer application'. This is the consequence of a new paradigm in heavy flavour physics where {\em high precision} is added to {\em high sensitivity}. 
Many questions have been raised and problems been suggested about the experimental reach of such a machine; answering them requires highly nontrivial work. A straightforward answer can be only a negative one, i.e. that such a challenging experimental set cannot be realized for technical or financial reasons. A positive answer, i.e. to go ahead with the construction of a Super-$B$ factory cannot be based on facts alone; it has to driven by our ambitions and a bold evaluation of our goals, i.e. a vision of what the field of high energy physics is all about.

\section*{Acknowledgments}
This paper represents a somewhat extended version of two talks given at the 5th Workshop on a 
Higher Luminosity $B$ Factory, Sept. 24 - 26, 2003, at Izu, Japan. We thank the organizers led by 
Masanori Yamauchi for finding such a splendid site and creating a truly inspiring atmosphere. 
This work has been supported by the NSF under the grant PHY-0087419 and by the US-Japan Cooperative 
Science Program under the grant INT-0089550. The work of AIS was also supported in part by Grant-in-Aid for 
Scientific Research from the Ministry of Education, Science and Culture of Japan.

\end{document}